\documentclass[conference]{IEEEtran}
\IEEEoverridecommandlockouts
\usepackage{cite}
\usepackage{amsmath,amssymb,amsfonts}
\usepackage{algorithmic}
\usepackage{graphicx}
\usepackage{textcomp}
\usepackage{xcolor}
\usepackage{amsthm, algorithm, float, geometry, breakcites, enumitem, url, booktabs, arydshln}
\usepackage{mathrsfs}

\def\BibTeX{{\rm B\kern-.05em{\sc i\kern-.025em b}\kern-.08em
    T\kern-.1667em\lower.7ex\hbox{E}\kern-.125emX}}
\begin{document}

\title{Physics-Constrained Backdoor Attacks on Power System Fault Localization}

\author{
Jianing Bai, Ren Wang, Zuyi Li
\thanks{Ren Wang and Zuyi Li are with the Department of Electrical and Computer Engineering, Illinois Institute of Technology, Chicago, IL, US. Emails: $\{\text{rwang74},\text{lizu}\}$@iit.edu. Jianing Bai is with the Department of Mechanical Engineering, Peking University, Beijing, China. Email: bai@stu.pku.edu.cn. This work was done when Jianing Bai was a research intern at the Trustworthy and Intelligent Machine Learning Research Group under the supervision of Ren Wang. (Corresponding author: Ren Wang)}}


\maketitle

\begin{abstract}
The advances in deep learning (DL) techniques have the potential to deliver transformative technological breakthroughs to numerous complex tasks in modern power systems that suffer from increasing uncertainty and nonlinearity. However, the vulnerability of DL has yet to be thoroughly explored in power system tasks under various physical constraints. This work, for the first time, proposes a novel physics-constrained backdoor poisoning attack, which embeds the undetectable attack signal into the learned model and only performs the attack when it encounters the corresponding signal. The paper illustrates the proposed attack on the real-time fault line localization application. Furthermore, the simulation results on the 68-bus power system demonstrate that DL-based fault line localization methods are not robust to our proposed attack, indicating that backdoor poisoning attacks pose real threats to DL implementations in power systems. The proposed attack pipeline can be easily generalized to other power system tasks.
\end{abstract}


\section{Introduction}
The modern power system has displayed surprising behaviors due to the high uncertainty brought on by renewable energy, the high nonlinearity resulting from the interconnection of power grids, and the high diversity of data collected by various sensors. Deep learning (DL)-based approaches, in contrast to many traditional approaches that struggle to manage systems with increasing complexity, provide promising solutions for complex problems like load and power forecasting \cite{wang2021review} 
and stability control \cite{alimi2020review}. Furthermore, fault detection and localization, our primary studying object, plays an essential role in the operations of electric grids and has also been advanced by DL methods \cite{li2019real,guo2019deep,fahim2021deep}. However, the introduction of DL raises new robustness concerns beyond conventional threads like the false data injection attack (FDIA), which aims to manipulate sensor measurements to perturb the results of power system state estimation without being detected \cite{zheng2021algorithm}. 

Deep neural networks (DNNs), stacked by multiple layers and can identify underlying relationships in a piece of data, are the foundation of deep learning. Recent works in the computer vision domain have demonstrated DNNs' vulnerability when facing training-phase backdoor poisoning attacks \cite{YLZ19,gu2019badnets}, and inference-phase adversarial attacks \cite{carlini2017towards,wang2022ask}. When considering power system tasks with physical constraints, there remain substantial difficulties in designing success attacks as these attacks should simultaneously achieve a high attack success rate and satisfy physical constraints to bypass detection methods. The vulnerability of post-trained DNNs against inference-phase adversarial attacks has attracted a great deal of attention in power system tasks \cite{tian2021joint,chen2019exploiting,chen2018machine}. A more stealthy and harmful attack type, the backdoor poisoning attack (a.k.a. Trojan attack), happens during the training phase and could cause erroneous behavior of DNNs when polluting a small portion of training data. DL systems in downstream applications could suffer severe damage if DNNs are not robust against backdoor assaults, raising serious questions about their reliability. To our best knowledge, no work has considered training-phase backdoor poisoning attacks in power systems (PSs).

The risk of backdoor attacks on fault line localization techniques based on DL is examined in this work. Specifically, we take the power systems' structures and physical laws into account to design the training-phase backdoor poisoning attacks. 
We summarize our contributions below:
\begin{itemize}
\item We design a novel physics-constrained backdoor attack strategy on DL-based fault line localization tasks in power systems.
\item We consider different threat models where attackers can directly manipulate training data or only access measurements.
\item By conducting fault localization simulations on the IEEE 68-bus power system, we demonstrate that the proposed physics-constrained backdoor attacks have the power to fail the predictions with the pre-designed triggers by using just a small number of poisoning training data while still maintaining a high accuracy on clean data.
\end{itemize}

\section{Preliminaries}\label{Pre}
\subsection{Power Grids and Fault Localization}
The topology of power grids can be abstracted as networks $G(\mathcal{N}, \mathcal{E})$ that include two main components: $\mathcal{N}$ buses (nodes) and $\mathcal{E}$ transmission lines (edges) that connect these buses. For an $d$-bus power grid, before the fault happens, the bus voltages ${\bf u}^{0} \in \mathbb{C}^{d} $, currents ${\bf i}^{0} \in \mathbb{C}^{d} $, admittance matrix $Y^{0} \in \mathbb{C}^{d \times d}$  follow the Ohm's law
\begin{eqnarray}
{\bf i}^{0}=Y^{0}{\bf u}^{0}
\label{eq: Ohm1}
\end{eqnarray}
where the entry of $Y^{0}$ is $y^{0}_{ij}$, denoting the admittance between the bus $i$ and $j$. Analogously, when the fault occurs, the bus voltages  ${\bf u^{'}} \in \mathbb{C}^{d} $ and the currents ${\bf i^{'}} \in \mathbb{C}^{d} $ also obey Ohm's law. We recommend readers to \cite{li2019real} for more details.

Fault localization, which aims to predict the faulted line, is the selected application in this work. Although there are various types of methods to predict fault location in power systems, DL-based methods leveraging features of currents provide state-of-art results \cite{li2019real, zhang2020novel}. The current varies with the fault position in real-time and has relatively larger fluctuations. Therefore using current can result in better performances compared with other signals.

\subsection{Deep Neural Networks}
During the DL training, all parameters of DNNs are optimized to minimize a loss function for increasing the prediction probabilities of ground truth classes.
Different from the fully-connected neural networks (FCNN) that have weights connections among all nodes, convolutional neural networks (CNNs) have shared weights and the ability of local feature exaction \cite{yoon2022deep}. 

\section{Physics-Constrained Backdoor Attacks}\label{PCBA}

\subsection{Threat Models}
The adversary aims to achieve a high attack success rate on modified inputs and high clean accuracy on original inputs. We consider two scenarios. In the first scenario, we assume that the adversary can directly manipulate the training set. In the second scenario, the adversary can only change sensor measurements corresponding to active and reactive power or voltage magnitude and phase angle. In both cases, the adversary is allowed to change labels fault localization as they are the direct observations.

\subsection{Problem Formulation}

Adversary aims to inject some pre-designed patterns into a small portion of training data to affect downstream tasks. In the fault localization, an adversary injects a signal (backdoor trigger) to training examples, resulting in the post-trained model predicting a pre-assigned electric power line when it sees the signal in inputs. Mathematically, given the training dataset with data feature ${X} \in \mathbb{R}^{n \times d}$ and the label ${Y} \in \mathbb{R}^{n}$ of size $n$ and dimension $d$, the neural network training process under backdoor attack is to solve the optimization problem below
\begin{equation}
\begin{aligned}
\min_{\boldsymbol \theta}& \mathcal L({\boldsymbol \theta};(X_{b},Y_{t})\cup (X_{2},Y_{2})), \\&
s.t. \quad X_{b}=h(X_{1};\Omega ; \mathcal C),
\label{eq: nn backdoor1}
\end{aligned}
\end{equation}
where $\mathcal L$ denotes the loss function, which is the cross-entropy in our setting. $\boldsymbol \theta$ denotes neural network parameters. ${X_1} \in \mathbb{R}^{r \times d}$ and ${X_2} \in \mathbb{R}^{(n-r) \times d}$ are two non-overlapped subsets of $X = \begin{bmatrix} X_1 \\ X_2 \end{bmatrix} \in \mathbb{R}^{n \times d}$. ${Y_1} \in \mathbb{R}^{r}$ and ${Y_2} \in \mathbb{R}^{n-r}$ are label subsets corresponding to $X_1$ and $X_2$, respectively. In the fault localization task, labels include $m$ different locations and one normal condition. 
${Y_{t}} \in \mathbb{R}^{r}$ is the label set that replaces labels in $Y_1$ with the predetermined target label $y_t$, i.e., a fixed location or the normal condition. $h(\cdot)$ is a trigger injection mapping from $X_1$ to poisoned data ${X_{b}} \in \mathbb{R}^{r \times d}$ following the policy $\Omega$ and some physical constraints in the constraint set $\mathcal C$. After the training, $\boldsymbol \theta$ will predict $h({\mathbf x};\Omega; \mathcal C)$ to the target label $y_t$ for any input $\mathbf x$.

There are many choices for $X$ in the fault localization setting. Here we will follow the same line of \cite{li2019real}, in which the feature vector ${\boldsymbol \psi_q} \in \mathbb{C}^{d}$ (unit of current) shown in \eqref{eq: feature1} is used for $X$.
\begin{equation}
\begin{aligned}
&{\boldsymbol \psi=\boldsymbol \psi_{p}+j \boldsymbol \psi_{q}}=Y^{0} \Delta {\bf u},\\&
\boldsymbol \psi_{q}=Y_{p}^{0}\Delta {\bf u_{q}}+Y_{q}^{0}  \Delta {\bf u_{p}},
\label{eq: feature1}
\end{aligned}
\end{equation}
where $\Delta {\bf u = (u^{'}-u^{0}})\in \mathbb{C}^{d}$. Variable notations with subscripts $p$ and $q$ (e.g., $\boldsymbol \psi_{p}$ and $\boldsymbol \psi_{q}$) denote the real part and the imaginary part of the original variable (e.g., $\boldsymbol \psi$).   

Note that in our setting, the poisoned data generated from mapping $h(\cdot)$ are physics-constrained by $\mathcal C$ to guarantee the effectiveness and practicality of the attack in power systems. Besides the Ohm's law constraint we introduced in \eqref{eq: feature1}, some typical constraints of $\mathcal C$ in the power system domain are listed as follows: 

\paragraph{Power Flow Constraints}
The power flow constraints are the mapping $h^{p}(\cdot)$ and $h^{q}(\cdot)$ from the voltage magnitude $\bf v$ and phase angle $\boldsymbol \theta$ to the real power $\bf p$ and the reactive power $\bf q$. 
\begin{equation}
{\bf p}=h^{p}({\bf v},\boldsymbol \theta),{\bf q}=h^{q}({\bf v},\boldsymbol \theta),
\label{eq: constraints}
\end{equation}

\paragraph{Power Limit Constraints}
The power, voltage, and current flow at all points on the system must be maintained within equipment operating limits to prevent damage to equipment. 
\begin{equation}
\label{eq: limits}
{\bf \underline{g}}\le {\bf g}\le {\bf \overline{g}},
\end{equation}
where ${\bf g}$ is a general notation for line flows, generations, voltage magnitudes, and phase angles. ${\bf \underline{g}}$ and ${\bf \overline{g}}$ are ${\bf g}$'s element-wise lower and upper bounds.

\paragraph{Bad Data Detection Under State Estimation}
$\bf v$ and $\boldsymbol \theta$ are usually estimated from $\bf p$ and $\bf q$ according to the measurement Jacobian matrix $H$. Therefore the perturbations $\delta ^{\bf v,\boldsymbol \theta }$ on $\bf v,\boldsymbol \theta$ and the perturbations $\delta ^{{\bf p},{\bf q}}$ on ${\bf p},{\bf q}$ should also obey the following equation.
\begin{eqnarray}
\delta ^{{\bf p},{\bf q}}=H\delta ^{\bf v,\boldsymbol \theta },
\label{eq: SE}
\end{eqnarray}

We will next show how to generate the backdoor counterpart of $\boldsymbol \psi_{q}$ when the adversary has different knowledge.

\subsection{Backdoor Trigger Design}

Given clean input feature $\boldsymbol \psi_{q} \in \mathbb{R}^{d}$, the backdoor perturbation $\Delta{\boldsymbol \psi_{q}}$ can be generated by
\begin{align}
    \begin{array}{ll}
\Delta{\boldsymbol \psi_{q}} &=\boldsymbol \psi_{q}^\prime - \boldsymbol \psi_{q}=  ({\bf 1_{d} - {\bf m})} \cdot \boldsymbol \psi_{q} + \bf m \cdot \boldsymbol \delta - \boldsymbol \psi_{q} 
\\ &= 
{\bf m} \cdot \boldsymbol \delta - {\bf m} \cdot (Y_{p}^{0} \Delta {\bf u_{q}}+Y_{q}^{0} \Delta {\bf u_{p}})
    \end{array}
    \label{eq: backdoor_new}
\end{align}
and constrained by $\mathcal C$, where the backdoor data $\boldsymbol \psi_{q}^\prime$ is encoded by the binary mask  ${\bf m} \in \{0,1\}^d$ and the element-wise perturbation $\boldsymbol \delta \in \mathbb{R}^{d}$. ${\bf m}$ and $\boldsymbol \delta$ decide the backdoor position and magnitude, respectively. $\bf 1_{d}$ represents all one vector with dimension $d$. $\cdot$ is the element-wise product. 

After changing the labels corresponding to the modified data samples to the target label, the classifier parameter $\boldsymbol \theta$ is trained based on \eqref{eq: nn backdoor1}, in which poisoned data-label samples are injected. The injected backdoor signals do not affect the model's behavior on clean inputs but will force the model to predict the target label if we add the trigger $({\bf m},\boldsymbol \delta)$ to an input in the inference phase. However, in most cases, the adversary cannot directly change $\boldsymbol \psi_{q}$ but can only manipulate the sensor measurements, which are voltage or complex power. In what follows, we show how to manipulate sensor measurements to lead to effective backdoor perturbation on training data.

\begin{figure*}[ht]
	\begin{center}
		\centering
	\includegraphics[width=0.89\linewidth]{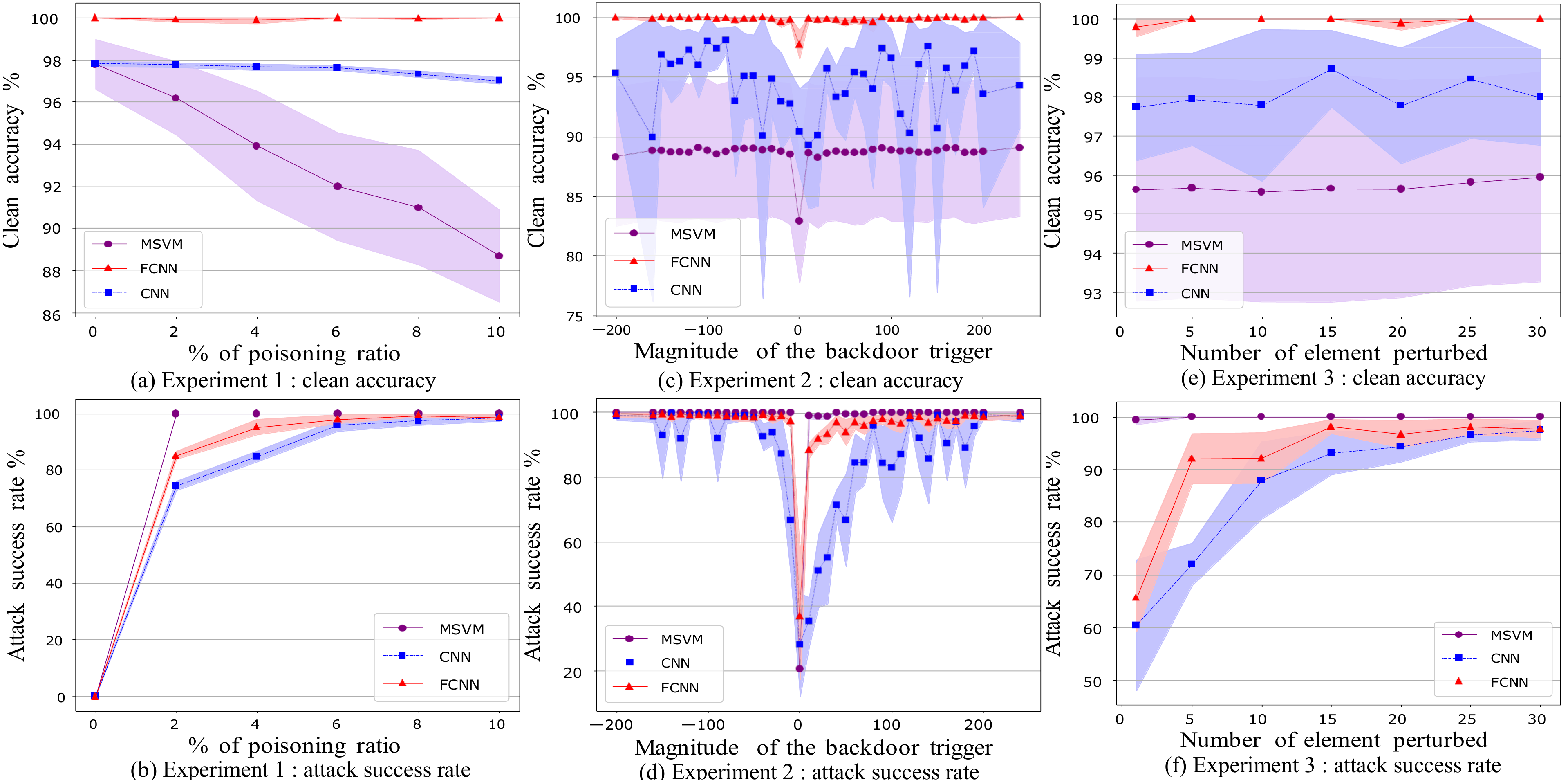}
		\caption{\textbf{Experiment (1):} With \textbf{only one non-zero entry in the backdoor trigger and backdoor magnitude 150}, the proposed attack achieves high success rates (predicting most of the test data to location 1 of the 86 locations) and clean accuracy as the poisoning ratio varies from 2\% - 10\%. \textbf{Experiment (2):} With \textbf{only one non-zero entry in the backdoor trigger and poisoning ratio 10\%}, the attack success rate increases when the absolute value of magnitude increases; \textbf{Experiment (3):} With the \textbf{1\% poisoning ratio and backdoor magnitude 50}, the attack success rate increases and the clean accuracy maintains when the number of non-zero entries in the trigger increases. (Each point is an average of 14 trials; The shadows are 95\% confidence intervals)
		}
		\label{comparison1}
	\end{center}
 \vspace{-8mm}
\end{figure*}

If an adversary can only access power data $\bf s$ consisting of $\bf p$ and $\bf q$, the attacker needs to obtain the desired backdoor perturbation $\Delta {\boldsymbol \psi_{q}}$ through the chain $\Delta \bf s \rightarrow {\Delta^2{\bf u}}\rightarrow \Delta{\boldsymbol \psi_{q}}$, where $\Delta \bf s$ and ${\Delta^2{\bf u}}$ are the perturbations of power $\bf s$ and voltage difference ${\Delta{\bf u}}$, respectively. The adversary first estimates $\bf u$ and $\bf u^{'}$ via state estimation and forces the estimation to satisfy the power flow and power limit. Then one can deduce the relationship between ${\Delta^2{\bf u}}$ with $\Delta {\boldsymbol \psi_{q}}$ according to \eqref{eq: feature1}, represented by $\Delta {\boldsymbol \psi_{q}}=f_{1}({\Delta^2{\bf u}})$. Note that we only need to perturb one of $\bf u_{p}$ or $\bf u_{q}$, where $\bf u$ can be $\bf u^{0}$ or $\bf u^{'}$. Finally, we can characterize the relationship between ${\Delta^2{\bf u}}$ and $\Delta \bf s$ based on \eqref{eq: SE}. The modified voltage and power also need to satisfy \eqref{eq: constraints} and \eqref{eq: limits}. The mapping from $\Delta \bf s$ to ${\Delta^2{\bf u}}$ is represented by ${\Delta^2{\bf u}}=f_2(\Delta \bf s)$. $\Delta {\boldsymbol \psi_{q}}$ can be injected following the mapping $f_1(f_2(\cdot))$.

We remark that the freedom of creating $\Delta \bf s$ and ${\Delta^2{\bf u}}$ is high because $\bf m$ is usually sparse, resulting in sparse $\Delta {\boldsymbol \psi_{q}}$. For example, we select $\bf m$ with only one non-zero entry in most of our experiments.

\section{Experimental Results}\label{ESR}

\subsection{Experimental Settings}

The dataset used in our experiments includes 1642 data samples (80\% as training data and 20\% as test data) obtained by simulating the IEEE 68-bus power system through Power System Toolbox (PST) \cite{li2019real}. The dataset is labeled into $87$ classes, in which the first $86$ classes correspond to the location of the faulted line and the $87$th class represents the normal condition. The dataset includes four types of faults: three-phase short circuit (TP), line-to-ground (LG), double line-to-ground (DLG), and line-to-line (LL). Our baseline network for this task consists of three types of classifiers, including multiple support vector machine (MSVM) \cite{poyhonen2005coupling}, three-layer fully-connected neural network (FCNN) \cite{lecun2015deep}, and a CNN with four convolutional layers and one fully connected layer \cite{krizhevsky2012imagenet}. The target label is set to be location 1 by default. The poisoning ratio is the percentage of poisoning data in the training set.

\subsection{Attack Results}

Fig.~\ref{comparison1} (a) and (b) demonstrate the effectiveness of the proposed attack. By choosing 150 as the backdoor magnitude and \textbf{only perturbing the first element of inputs}, we can find that the attack success rates increase and the clean accuracy remain at a similar level for neural networks as the poisoning ratio increases from 0\% to 10\%. \textbf{A high attack success rate means that the attack has forced the model to predict most of the poisoned test data to location 1, which will potentially lead to increased system misoperation and maybe even cause severe cascading failures or blackouts.}
At the point of the 10\% poisoning ratio, all three poisoned models' average attack success rates are higher than 98.40\%, while the clean accuracy drops are 0.04\%/1.81\%/9.31\% for FCNN/CNN/MSVM compared with the 0\% poisoning ratio scenario. The 9.31\% accuracy drop of the MSVM is probably a result of its weak generalization ability.

Fig.~\ref{comparison1} (c) and (d) show the impact of the magnitude of the backdoor trigger on clean accuracy and attack success rate, respectively. By injecting 10\% poisoned samples into the training dataset and \textbf{randomly perturbing one element}, we can find that the attack success rates and clean accuracy are both affected at points around zero. With the increase of the absolute value of the magnitude, the clean accuracy is almost unchanged, and the attack success rate rises rapidly. 

We also study the impact of the number of non-zero entries of backdoor triggers on clean accuracy and attack success rate. As shown in Fig.~\ref{comparison1} (e) and (f), under \textbf{1\% poisoned samples and backdoor magnitude 50}, we can find that with the increase of the number of non-zero entries (randomly selected), the clean accuracy is almost unchanged, and the attack success rate rises rapidly on FCNN and CNN. 

\section{Conclusion}\label{Concl} 

For the first time, we proposed a novel physics-constrained backdoor attack for evaluating the security of deep learning-based power system applications. The attack manipulates a small portion of training data points by injecting backdoor signals constrained by power system laws. In the inference phase, these backdoor signals can mislead the deep learning models to some target classes. The data manipulation can even happen on sensor measurements. It has been proven through simulations that our proposed attack can achieve high attack success rates and high clean accuracy simultaneously under various poisoning ratios. \textbf{Although this paper only considers the deep learning-based fault localization task, it can be naturally generalized to other applications and learning frameworks.}

\bibliographystyle{ieeetr}
\bibliography{refs/ref_mr,refs/ref_new}

\vspace{12pt}

\end{document}